# BOHR QUANTUM THEORY OF THE MAGNETIC MONOPOLES AND CLASSICAL ELECTRON ELECTROMAGNETIC MASS PROBLEM

Vladan Panković

Department of Physics, Faculty of Sciences, 21000 Novi Sad, Trg Dositeja Obradovića 4., Serbia

**Abstract**

In the first part of this work we apply Bohr (old or naïve quantum atomic) theory for analysis of the remarkable electro-dynamical problem of magnetic monopoles. We reproduce formally exactly some basic elements of the Dirac magnetic monopoles theory, especially Dirac electric/magnetic charge quantization condition. It follows after application of Bohr theory at the system, simply called magnetic monopole "atom", consisting of the practically standing, massive magnetic monopole as the "nucleus" and electron rotating stably around magnetic monopole under magnetic and electrostatic interactions. Also, in the second part of this work we suggest a simple solution of the classical electron electromagnetic mass problem.

## 1. Introduction

In the first part of this work we shall apply Bohr (old or naïve quantum atomic) theory for analysis of the remarkable electro-dynamical problem of magnetic monopoles [1]. We reproduce formally exactly some basic elements of the Dirac magnetic monopoles theory, especially Dirac electric/magnetic charge quantization condition. It follows after Bohr theory application at the system, simply called magnetic monopole "atom", consisting of the practically standing, massive magnetic monopole as the "nucleus" and electron rotating stably around magnetic monopole under magnetic and electrostatic interactions. Also, in the second part of this work we shall suggest a simple solution of the classical electron electromagnetic mass problem [2].

## 2. Bohr theory of the magnetic monopoles

As it is well known Dirac [1] introduced concept of the magnetic monopole starting, roughly speaking, from the rotation of the electron around magnetic monopole described by Dirac relativistic equation of the electron. It yields the following Dirac electric/magnetic charge quantization condition

(1) $\quad 1/(4\pi \varepsilon_0 c^2) \, eq = (n/2)\hbar$

for n=1, 2, … . Here e represents the electron electrical charge (absolute value), q - magnetic monopole magnetic charge, c – speed of light, $\varepsilon_0$ - vacuum electric permittivity correlated with vacuum magnetic permittivity $\mu_0 = 1/(\varepsilon_0 c^2)$, and $\hbar$ – reduced Planck constant.

According to Dirac theory distance between electron and magnetic monopole can be arbitrary. Dirac theory does not give any prediction on the magnetic monopole mass, but according to cotemporary quantum field theories it can be expected that this mass is much larger than electron mass.

So, consider the system, simply called magnetic monopole "atom", consisting of the practically standing, massive (with mass much larger than electron mass) magnetic monopole as the "nucleus" and electron rotating around this magnetic monopole.

Suppose that magnetic monopole holds magnetic charge q and corresponding electric charge q/c.

Suppose that under magnetic and electrostatic interactions electron rotates stably with speed v at distance R around resting magnetic monopole, so that the following condition of the circular orbit stability is satisfied

(2) $\quad \mu_0/(4\pi) \, (ev)q/R^2 + 1/(4\pi \varepsilon_0) \, e(q/c)/R^2 = mv^2/R$ .

Here first term at the left hand of (2) represents the classical attractive magnetic force between monopole and small system, second term at the left hand of (2) – attractive electrostatic force between monopole and small system, while right hand of (2) represents the amplitude of the centrifugal force for electron mass m.

Suppose additionally that magnetic and electrostatic forces have the same intensities, i.e.

(3) $\quad \mu_0/(4\pi) \, (ev)q/R^2 = 1/(4\pi \varepsilon_0) \, e(q/c)/R^2$ .

Then (1) turns out in

(4) $\quad 2 \mu_0/(4\pi) \, (ev)q/R^2 = mv^2/R$

which implies

(5) $\quad 1/(4\pi \varepsilon_0 c^2) \, e \, q = (1/2) mvR$ .

Finally, suppose that there is Bohr (old or naïve quantum theoretical) quantization of the angular momentum of the small system by rotation around magnetic monopole, i.e.

(6) $\quad mvR = n\hbar$

for n=1, 2, … , where $\hbar$ represents the reduced Planck constant.

Then, (5) turns out in

(7) $\quad 1/(4\pi \varepsilon_0 c^2) \, eq = (n/2) \, \hbar$

for n=1, 2, … .

As it is not hard to see expression (7) has form exactly equivalent to remarkable Dirac electric/magnetic charge quantization relation (1).

It can be observed that condition of the equivalence between magnetic and electrostatic force (3) implies

(8)    $v = c$

for n=1, 2, … . It means that electron at any circular orbit propagates with speed of light.

Also, introduction of (8) in (6) implies

(9)    $R = n \hbar/(mc) = n \lambda_{c\,red}$

for n=1, 2, … where $\lambda_{c\,red} = \hbar/(mc)$ represents the reduced Compton wavelength of the electron.

### 3. A simple solution of the electron electromagnetic mass problem

Remarkable problem of the electron electromagnetic mass in the classical electrodynamics Feynman [2] presented in the following way.

Consider firstly single electron in rest. Suppose that the charge of the electron e is continuously homogeneously distributed over surface of a sphere with some constant radius $a_0$. Energy density of the electrical field corresponding to such charge distribution according to classical electrodynamics equals

(10)    $u = 1/2\, \varepsilon_0 E^2 = 1/2\, e^2/(16\pi^2\, \varepsilon_0\, r^4) = e^2/(32\pi^2\, \varepsilon_0\, r^4)$

where $E = e/(4\pi\varepsilon_0 r^2)$ represents electrical field intensity at the distance r in respect to the sphere center for r larger than $a_0$.

Then total energy of the electric field can be obtained by simple integration of the (10) in the distance domain $[a_0, \infty)$ for the following differential volume element $4\pi r^2\, dr$, and it equals

(11)    $U = \int u\, (4\pi r^2\, dr) = 1/2\, e^2/(4\pi\varepsilon_0\, a_0)$    .

Obviously this expression tends toward infinity when $a_0$ tends toward zero and it represents one serious problem of the classical electrodynamics.

However, it can be supposed that expression (11) must be equivalent to the relativistic electron rest energy $m_0 c^2$, i.e.

(12)    $U = m_0 c^2$

where $m_0$ represents the electron rest mass. It implies the following constant value for *a* (that cannot be arbitrarily diminished)

(13)    $a_0 = e^2/(m_0 c^2 8\pi\varepsilon_0) = 1/2\, e^2/(m_0 c^2 4\pi\varepsilon_0) = 1/2\, r_0$

where $r_0 = e^2/(m_0 c^2 4\pi\varepsilon_0)$ represents so-called electron classical radius. In this way there is no mentioned problematic (zero for $a_{00}$ and infinity for U) limit.

Nevertheless, by motion of the electron with a small, non-relativistic constant speed v, for the electromagnetic field directed under angle $\theta_0$ in respect to the electron motion direction, there is a classical electrodynamical field momentum density in the direction of the electron motion

(14)    $g \sin \theta = (v/c^2\, \varepsilon_0 E^2 \sin \theta) \sin \theta$    .

It, after simple integration over the distance domain $[a, \infty)$ and over the angle domain $[0,\pi]$ for the following differential volume element $2\pi r^2 \sin \theta\, dr\, d\theta$, yields the following total field momentum in the direction of the electron motion

(15)    $p = (4/3\, e^2/(8\pi\varepsilon_0\, a\, c^2))\, v = m_{em}\, v$

where expression 4/3 represents the value of the integration over angle $\theta$ [2] and where

(16)    $m_{em} = 4/3\, e^2/(8\pi\varepsilon_0\, a\, c^2)$

represents so-called electron electromagnetic mass.

Obviously, this electromagnetic mass tends toward infinity when electron sphere radius *a* tends toward zero what represents a serious problem for the classical electrodynamics.

Additionally it can be supposed

(17)    $a = a_0$

as it is usual case [2]. Then moving electron sphere radius is limited, using (12), (13), and electron electromagnetic mass (16) cannot tend toward infinity. But then expression (16) yields

(18)    $m_{em} = 4/3 \, m_0$

that simply leads toward explicit contradiction with relativistic equivalence between mass and energy [2]. It can be ad hoc supposed that this contradiction can be solved by introduction of the additional, non-electromagnetic electron mass part caused by some hypothetical non-electromagnetic forces, i.e. so-called "Poincare stresses". But, really, there is no any completely consistent and complete model of such forces [2]. It represents well-known, open problem of the electron electromagnetic mass in the classical electrodynamics.

We shall suggest a very simple solution of this problem.

Firstly, we shall present mentioned problem more explicitly. By mentioned slow, non-relativistic movement of the electron, electromagnetic field energy density holds the following form

(19)    $u = 1/2 \, \varepsilon_0 E^2 + 1/2 \, \varepsilon_0 c^2 B^2 = 1/2 \, \varepsilon_0 \, (e/(4\pi\varepsilon_0 r^2))^2 + 1/2 \, \varepsilon_0 c^2 (v/c^2 \, e/(4\pi\varepsilon_0 r^2) \sin\theta)^2$
        $= e^2/(32\pi^2 \varepsilon_0 r^4) + e^2/(32\pi^2 \varepsilon_0 r^4) \, (v^2/c^2 \sin^2\theta)$

where $B = v/c^2 \, E \sin\theta$ represents corresponding magnetic field intensity. Expression (19), after simple integration over the distance domain $[a, \infty)$ and over the angle domain $[0,\pi]$ for the following differential volume element $2\pi r^2 \sin\theta \, dr \, d\theta$, yields the following total energy of the electric field

(20)    $U = \int u \, (2\pi r^2 \sin\theta \, dr \, d\theta) = \int e^2/(32\pi^2 \varepsilon_0 r^4)) \, (2\pi r^2 \sin\theta \, dr \, d\theta)$
        $+ \int e^2/(32\pi^2 \varepsilon_0 r^4)) \, (v^2/c^2 \sin^2\theta) \, (2\pi r^2 \sin\theta \, dr \, d\theta)$
        $= 2e^2/(16\pi\varepsilon_0 a)) + e^2/(16\pi\varepsilon_0 a)) \, (v^2/c^2 \, 4/3)$
        $= e^2/(8\pi\varepsilon_0 a)) + 1/2 \, e^2/(8\pi\varepsilon_0 a)) \, (v^2/c^2 \, 4/3)$ .

It can be observed that (20) is somewhat similar to the relativistic energy-mass equivalence form in non-relativistic approximation, i.e. to

(21)    $E = m c^2 = m_0 c^2/(1 - v^2/c^2)^{1/2} \approx m_0 c^2 (1 + 1/2 \, v^2/c^2) = m_0 c^2 + 1/2 \, m_0 c^2 \, v^2/c^2 =$
        $= m_0 c^2 + 1/2 \, m_0 v^2$    .

Namely, under supposition (17) and according to (13), expression (18) turns out in

(22)    $U = m_0 c^2 + (4/3) \, 1/2 \, m_0 v^2$

that is similar but not identical to (21). It, of course, represents mentioned electron electromagnetic mass problem.

For a simple solution of this problem we shall suppose instead of (17), (18) the following. Concretely, we shall suppose

(23)    $m_{em} = m_0$

, or, we shall suppose that there is no any difference between electron rest and electromagnetic mass (if this supposition is correct there is no any problem with the electron electromagnetic mass). But then, according to (16) and (13) it follows

(24)    $a = 4/3 \, a_0$

which means that the moving electron sphere radius is different than rest electron sphere radius.

Further, in the expression for the electromagnetic field energy density (19) first term refers on the static and the second term on the moving electron. For this reason, total energy of the electromagnetic field can be obtained by integration of (19), but mentioned first term of (19) must be integrated over distance domain $[a_0, \infty)$ for the following differential volume element $4\pi r^2\, dr$, while second term of (19) must be integrated over the distance domain $[a, \infty)$ and over the angle domain $[0, \pi]$ for the following differential volume element $2\pi r^2 \sin\theta\, dr\, d\theta$. Then, as the final result, we obtain

(25)    $U = e^2/(8\pi\varepsilon_0\, a_0)) + 1/2\ e^2/(8\pi\varepsilon_0\, a))\ (v^2/c^2\ 4/3)$ .

It, according to (13) and (24) turns out in

(26)    $U = m_0 c^2 + 1/2\ m_0 v^2$

identical to relativistic expression (21).

In this way electron electromagnetic mass is simply and consistently solved.

### 4. Conclusion

In conclusion we can shortly repeat and point out the following. In the first part of this work we apply Bohr (old or naïve quantum atomic) theory for analysis of the remarkable electro-dynamical problem of magnetic monopoles. We reproduce formally exactly some basic elements of the Dirac magnetic monopoles theory, especially Dirac electric/magnetic charge quantization condition. It follows after application of Bohr theory at the system, simply called magnetic monopole "atom", consisting of the practically standing, massive magnetic monopole as the "nucleus" and electron rotating stablely around magnetic monopole under magnetic and electrostatic interactions. Also, in the second part of this work we suggest a simple solution of the classical electron electromagnetic mass problem.